\documentclass[preprintnumbers,amsmath,amssymb]{revtex4}

\usepackage{graphicx}

\newcounter{myctr}
\def\myitem{\refstepcounter{myctr}\bibfont\noindent\ifnum\themyctr>9\else\phantom{0}\fi\hangindent17pt\themyctr.\enskip}

\def\ket#1{\ensuremath{{|#1\rangle}}}
\newcommand{\bra}[1]{\ensuremath{\langle#1\vert}}
\newcommand{\ketbra}[1]{\ket{#1}\bra{#1}}

\def\Cl{\overline{QC}}

\def\Kl{\overline{QK}}

\newcommand{\txt}[1]{\textrm{#1}}
\def\lbar{\ensuremath{\overline{l}}}
\def\ensemble{{\cal E}}
\def\mixture{\ensuremath{{\cal E}} }
\def\mixturedef{\ensuremath{{\cal E}=\{(p_i,\ket{\psi_i})\}} }

\def\be{\begin{eqnarray}}
\def\ee{\end{eqnarray}}

\newcommand{\nix}{{\rule{0pt}{2pt}}}
\newcommand{\qedd}{{\nix\nolinebreak\hfill\hfill\nolinebreak$\Box$}}
\newcommand{\qed}{{\qedd\par\medskip\noindent}}

\newtheorem{theorem}{Theorem}[section]

\newtheorem{lemma}[theorem]{Lemma}
\newtheorem{definition}[theorem]{Definition}

\newtheorem{example}[theorem]{Example}

\usepackage{color}
\usepackage{times}
\usepackage{float}
\usepackage{graphics}

\begin{document}
\title{Second Quantized Kolmogorov Complexity}			

\author{Caroline Rogers}

\address{Department of Computer Science, University of Warwick, Coventry CV4 7AL, United Kingdom}

\author{Vlatko Vedral}
\address{The School of Physics and Astronomy, University of Leeds, Leeds, LS2 9JT, United Kingdom\\Centre for Quantum Technologies, National University of Singapore, 3 Science Drive 2, Singapore 117543\\Department of Physics, National University of Singapore, 2 Science Drive 3, Singapore 117542}

\author{Rajagopal Nagarajan}
\address{Department of Computer Science, University of Warwick, Coventry CV4 7AL, United Kingdom}

\begin{abstract}
The Kolmogorov complexity of a string is the length of its shortest description.
We define a second quantised Kolmogorov complexity where the length of a description
is defined to be the average length of its superposition.
We discuss this complexity's basic properties.
We define the corresponding prefix complexity
and show that the inequalities obeyed by this prefix complexity
are also obeyed by von Neumann entropy.
\end{abstract}
\maketitle

\section{Introduction}

Quantum physics is a more accurate description of physical
phenomena than classical mechanics. If we therefore wish to speak about the complexity of some
processes or some physical states in nature, it is more accurate to use a quantum mechanical model. An important
question is (in very simple terms), given a physical system in some state, how difficult
is it for us to reproduce it. If we wish to have a universal
measure of this difficulty (which applies to all systems and
states) a way to proceed is to follow the prescription of
Kolmogorov.

Kolmogorov complexity is a very general measure of information.
The Kolmogorov complexity of a state is defined as the shortest
description of that state on a universal computer. The intuition is that if a state
can be described very succinctly, then that state is very simple, whereas
complex states have very long descriptions. Kolmogorov
complexity is a well-developed field with a plethora of applications in areas
such as computer science, physics, pure mathematics and philosophy \cite{vitanyi}.
In this paper, a second quantised Kolmogorov complexity based on average lengths of indeterminate length descriptions is defined. It is called second quantised because,
as we shall explain, the physical interpretation of this complexity
is of a system in the second quantisation.
Its basic properties
are then discussed; its physical interpretation, its incompressibility, the complexity of multiple
copies of a state. We then define the corresponding prefix complexity
and show that it is also incompressible, that it is related to von Neumann entropy,
and that any inequality obeyed by the prefix complexity is also obeyed
by von Neumann entropy.

\section{Synopsis}

Section III describes the basic properties of classical Kolmogorov complexity. In Section IV, the related previous work
in quantum information theory is described.
In Section V, the second quantised Kolmogorov complexity based on
average length descriptions is defined and its basic properties are discussed. In Section VI, the corresponding prefix Kolmogorov complexity is defined. It is shown that the complexity of a superposition is not additive and that the
expected complexity of a mixture is closely related to its von Neumann entropy.

\section{Classical Kolmogorov Complexity}

Kolmogorov complexity is an information complexity based on the intuition that a simple string has a short description.
The string $00000000000000000000$, which
 can be described as ``twenty $0$'s", seems much simpler than say $10010111011010100010$
in which there is no obvious pattern (though one might exist).
To avoid self-contradictory statements such as ``the shortest string that cannot be described in less than
one hundred words", Kolmogorov complexity is defined with respect to a Turing machine.

\begin{definition}[Kolmogorov complexity with respect to $T$]
The Kolmogorov complexity of a string $x$ can be defined with respect to a particular Turing machine $T$ as the
length of the shortest string $p$, which when inputted into $T$, outputs $x$.
\[
C_T(x)=\min_{T(p)=x}l(p)
\]
\end{definition}
Sometimes it is useful to be able to input two strings $x$ and $y$ into a classical Turing machine.
However, $x$ and $y$ cannot be obtained from the string $z=xy$ without encoding where the string $x$ ends.
Here is a simple way to encode a pair or a sequence of strings as input.
\begin{definition}[Encoding ordered sequences]
For any pair of binary strings $x_1$ and $x_2$, let
\[
(x_1, x_2) = 1^{l(x_1)}0x_1x_2
\]
(where $1^n$ means $n$ copies of $1$).
For any sequence of binary strings $x_1$, $x_2$, $\ldots$, $x_{m-1}$, $x_m$ let
\[(x_1,x_2, x_{m-1}, \ldots, x_m) = (x_1, (x_2, (\ldots(x_{m-1}, x_m) )))\]
\end{definition}
For example, $(110,1000) = 11101101000$. Reading $11101101000$ from left to right, the first three $1$s followed by a $0$ show that the first string has length $3$.
The next three bits give the value of the first string $110$, the remaining
four bits give the second string $1000$.

An essential feature of Kolmogorov complexity is that it does not really matter which Turing machine
is used to define it. The Kolmogorov complexity of a string $x$ is invariant in the
sense that there is a universal Turing machine $U$ such that for any Turing machine $T$
a constant exists which bounds $C_T(x)-C_U(x)$ for any $x$.
\begin{theorem}[Invariance theorem]\label{invariance}
There is a universal Turing machine $U$ such that for any Turing machine $T$ there exists a constant $c_T$
(dependent only on $T$ and $U$)
 such that for all $x$:
\[C_U(x) \leq C_T(x)+c_T\]
\end{theorem}
{\bf Proof.}
Each Turing machine has a finite set of instructions, hence the set of all Turing machines can be enumerated.
 Let $T_1$, $T_2$, $\ldots$ be
an enumeration of the Turing machines.
Let $U$ be a Turing machine which on input $(i,p)$, generates a description of the instructions of Turing machine $T_i$ and
simulates $T_i$ on input $p$ (i.e. $U(i,p)=T_i(p)$). If $p$ is a description of $x$ on Turing machine $T_i$, then
\[C_{U}(x)\leq l(p)+c_T\] where $c_T$ is a constant, independent of $x$,
dependent on the number of bits used to describe $i$.
\qed

Let us now fix the universal Turing machine $U$ in the invariance theorem, and define Kolmogorov complexity with respect to this reference machine $U$.
\begin{definition}[Kolmogorov complexity]
The Kolmogorov complexity of a string $x$ is defined as the shortest
description of $x$ with respect to the universal Turing machine:
\[
C(x)=\min_{U(p)=x}l(p)
\]
where $U$ is the fixed universal Turing machine as in the Invariance Theorem (Theorem \ref{invariance}).
\end{definition}
Sometimes it is also useful to define the complexity of $x$ conditional on knowing another string $y$. For example,
conditional complexity can be used to define how much information $y$ provides about $x$.
\begin{definition}[Conditional Kolmogorov complexity]
The conditional Kolmogorov complexity of a string $x$ given a string
$y$ is defined as
\[
C(x|y)=\min_{U(y,p)=x}l(p)
\]
where $U$ is the fixed universal Turing machine as in the Invariance Theorem (Theorem \ref{invariance}).
\end{definition}

\subsection{Incompressibility}

Many ``long strings" have short descriptions. For example, there is a Turing machine $T$
that outputs $2^{2^{2^n}}$ on input $n$, so for all $n$, $2^{2^{2^n}}$ can be described
using $\log(n)+c$ bits. One might hope that many large strings can be described by short strings.
However, the incompressibility theorem shows that not all strings are compressible.
\begin{theorem}[Incompressibility theorem]\label{incompressibility}
For any integer $n>0$, there is at least one string $x$ of length $l(x)=n$ which has a Kolmogorov complexity
is at least $n$.
\end{theorem}
{\bf Proof.}
The proof of the incompressibility theorem is a simple pigeon hole argument.
There are $2^n$ strings of length $n$ but only $2^0 + \ldots + 2^{n-1}<2^{n}$ strings of length less than $n$.
Therefore at least one string of length $n$ cannot be compressed to a string of length less than $n$.
\qed

\subsection{Prefix Kolmogorov Complexity}

The standard Kolmogorov complexity
$C$ which has been described so far does not deal with prefix free strings. Kolmogorov complexity can be related simply
to Shannon entropy by defining a prefix Kolmogorov complexity $K$ where the descriptions are prefix free.

Another advantage of prefix complexity over standard
complexity is that $K$ is additive, that is $K(x,y)\leq K(x)+K(y)+c$
(since $(x,y)$ can be described by concatenating prefix
free descriptions of $x$ and $y$). On the other hand, $C(x,y) \leq C(x)+C(y)+c$ does not hold in general \cite{vitanyi}. There are $(n+1)2^n$ pairs $(x,y)$ such that
$l(x)+l(y)=n$. Thus by incompressibility,  there exists at least one such
pair $(x,y)$ that has complexity
\be
C(x,y) & \geq &\log((n+1)2^n) \\
&=&n+\log(n)\\
&>&n+c\\
&=&l(x)+l(y)+c\\
&=&C(x)+C(y)
\ee

\subsubsection{Prefix Kolmogorov Complexity}

In order to define prefix Kolmogorov complexity, in which the descriptions are prefix free,
let us define a prefix
quantum Turing machine.
\begin{definition}[Prefix Turing machine]
$T$ is a prefix Turing machine if any two inputs $x$ and $y$ upon
which $T$ halts are prefix free.
\end{definition}
This definition is equivalent to restricting all inputs to be infinite length and defining
the length of the input to be the number of bits on the input tape that are used during computation.
A universal prefix Turing machine $U$ can be defined by enumerating the Turing machines $T_1$, $T_2$, $\ldots$
and using self-delimiting descriptions of each Turing machine. In this way, prefix Kolmogorov complexity
is invariant up to an additive constant.
\begin{definition}[Prefix Kolmogorov complexity]
The prefix Kolmogorov complexity $K$ of a string $x$ is:
\[K(x)=\min_{U(p)=x} l(p)\]
where $U$ is the reference universal prefix machine.
\end{definition}

\subsubsection{Prefix Kolmogorov Complexity and Shannon Entropy}

The expected prefix complexity of a random variable is closely related to the random variable's
Shannon entropy.
The expected prefix complexity of a random variable is defined as its average prefix complexity.
\begin{definition}[Expected prefix Kolmogorov complexity of a random variable]
Let $X$ be a random variable. The expected prefix Kolmogorov complexity of $X$ is:
\[E(K(X)) = \sum_x P(X=x)K(x)\]
\end{definition}
On the other hand, the complexity of a random variable $X$ is the complexity of describing
$X$ itself, given some suitable encoding.
\begin{definition}[Prefix Kolmogorov complexity of a random variable]
Given a random variable $X$, let $f(X)$ be an encoding of $X$ if
\[f(X)=((p_1,x_1),\ldots (p_m,x_m))\]
where $x_1\leq x_2 \leq \ldots \leq x_m$ and for all $x$ with $P(X=x)>0$, there exists a unique $1\leq i \leq m$ such that $x = x_i$ and $p_i=P(X=x)$.
The prefix Kolmogorov complexity of $X$ is the prefix Kolmogorov complexity of its encoding:
\[
K(X)=K(f(X))
\]
\end{definition}
Since a compression algorithm can be implemented by a Turing machine, $E(K(X))$ and $H(X)$ are closely
related.
\begin{lemma}[Relationship between $H$ and $K$]
There exists a constant $c$ such that for any random variable $X$,
\[
E(K(X)) - K(X) - c \leq H(X) \leq E(K(X))
\]
\end{lemma}
{\bf Proof.}
By Shannon's Noiseless Coding Theorem for Lossless Codes \cite{shannon}, there is no prefix free encoding
of $X$ whose expected length is less than $X$, thus $E(K(X))\geq H(X)$. On the other hand, by describing the random
variable $X$, a prefix code $F$ which encodes $X$ using $E(F(X)))=H(X)+1$ bits
can be described.
Thus $E(K(X)) \leq H(X)+K(X)+c$.
\qed


\section{Previous Work}

In this section, the previous attempts to define a quantum Kolmogorov complexity
and some of the previous work on indeterminate length quantum strings are described.
The first attempt to define a quantum Kolmogorov complexity was by Svozil \cite{svo96} in 1996.
The attempts to define a quantum Kolmogorov complexity \cite{ber01,gacs,mor05,segre,svo96,tad02,tad04,vit01} assume that the descriptions have determinate lengths.
Second quantised Kolmogorov complexity differs from the others in that it uses indeterminate length quantum strings.

\subsection{Indeterminate Length Quantum Strings}

Indeterminate length quantum strings have been discussed
in the context of lossless quantum data compression \cite{classicalsidechannels,bos02,losslessquantumcompression,sch01}. When a mixture of non-orthogonal strings is compressed
using a variable length string, the resulting compressed state may be in a superposition of different lengths.
Such a string is called an indeterminate length string \cite{sch01}.
\begin{definition}[Indeterminate length string]
$\ket{\psi}=\sum_i \alpha_i \ket{i}$ is an indeterminate length
quantum string if there exists $i$ and $j$ with $|\alpha_i|>0$ and
$|\alpha_j|>0$  and $l(i)\neq l(j)$.
\end{definition}
Determinate length strings of length $n$ exist in the Hilbert space $H^{\otimes n}$.
Indeterminate length strings exist in the Fock space
\[H^{\oplus}=\bigoplus_{n=0}^{\infty} H^{\otimes n} \]
The analysis of indeterminate length strings in the Fock space has recently been discussed \cite{markusprefix}. It was shown that an indeterminate
string in the Fock space $H^{\oplus}$ can be considered
as a string on a quantum Turing machine, where
a finite set of cells beginning from the first cell
are non-blank. The other cells contain the character \ket{\#} which is orthogonal to \ket{0} and \ket{1}.

Bostr\"{o}m and Felbinger \cite{bos02} defined two ways to quantify the lengths
of indeterminate length strings.
\begin{definition}[Lengths of indeterminate length strings]
The base length $L$ of an indeterminate length string is the length of the longest part of its superposition
\[L\left(\sum_i \alpha_i \ket{i}\right) = \max_{|\alpha_i|>0} l(i)\] The average length
$\lbar$ of an indeterminate length quantum string is the average length of its superposition
\[\lbar\left(\sum_i \alpha_i \ket{i}\right) = \sum_i |\alpha_i|^2 l(i)\]
\end{definition}
If the length of a quantum string is observed, then $\lbar$ is the expected length that is observed
and $L$ is the maximum length that can be observed.

\subsubsection{From Lossless Coding to Lossy Coding}

\label{lossyqcoding}
Schumacher and Westmoreland \cite{sch01} demonstrated
that by projecting onto
 $n(S(\ensemble)+\delta)$ qubits, if a mixture $\ensemble^{\otimes n}$
is encoded with a variable length condensable code, a fixed length lossy code can be obtained \cite{chu00}.
If $\ensemble$ is a mixture with density operator $\rho$, where $\rho$'s spectral
decomposition is:
\[\rho = \sum_i p_i \ketbra{i}\]
Then \mixture can be encoded by encoding each \ket{i} as a prefix free string
of length $\lceil-\log(p_i)\rceil$ with zero-padding.
$\rho^{\otimes n}$ can be encoded in the same
fashion. Almost every string \ket{i} in the typical subspace of $\rho^{\otimes n}$ has probability
arbitrarily close to $2^{-nS(\rho)}$ as $n$ grows large. Thus almost every string in the typical subspace of $\rho$
is encoded as a string of length arbitrarily close to $nS(\rho)$. By projecting onto $n(S(\rho)+\delta)$ qubits, the compressed state is projected onto the encoded
typical subspace of $\rho$. The typical subspace can be decoded to obtain the original mixture
\mixture with arbitrarily high (but not perfect) probability and fidelity.
Thus the minimum expected average length of a variable length code is related to von Neumann entropy.

From this encoding, it can be seen that the average lengths of  condensable codes obey Kraft's inequality (if they did not, then
a mixture could be lossily compressed to less than its von Neumann entropy).
\begin{lemma}[Kraft's inequality for condensable strings]
If $\xi$ is a set of orthogonal condensable strings then
\[
\sum_{\ket{\psi}\in \xi} 2^{-\lbar(\ket{\psi})} \leq 1
\]
\end{lemma}
{\bf Proof.}
Let $\xi = \{\ket{\psi_1}, \ldots, \ket{\psi_n}\}$ be a set of orthogonal condensable strings
and let
\[
c = \sum_{i=1}^n 2^{-\lbar(\ket{\psi_i})}
\]
Let $\rho$ be density operator defined by:
\[
\rho = \frac{\sum_{i=1}^n 2^{-\lbar(\ket{\psi_i})} \ketbra{\psi_i}}{c}
\]
Then the expected average length of $\rho$ is:
\be
E(\lbar(\rho)) &=&  \frac{\sum_{i=1}^n 2^{-\lbar(\ket{\psi_i})} \lbar(\ket{\psi_i})}{c} \\
&=& \frac{S(\rho)}{c}\\
\ee
As discussed above, it is possible to
design a lossy quantum code which has rate of compression $E(\lbar(\rho))$.
If $c>1$ then $E(\lbar(\rho)) < S(\rho)$. Therefore $c\leq 1$ as required.
\qed

\subsection{Quantum Kolmogorov Complexity}

Berthiaume, van Dam and Laplante \cite{ber01} defined a quantum Kolmogorov complexity based on Bernstein and Vazirani's
model of a quantum Turing machine \cite{ber97}. This quantum Kolmogorov complexity is invariant up to an
additive constant with respect to a universal quantum
Turing machine $U$ which can simulate any other quantum Turing
machine from its classical description. The quantum Kolmogorov
complexity $QC$ was defined as the base length of its shortest quantum
description on $U$
\[QC\ket{\psi} = \min_{U(p)=\ket{\psi}} L(p)\]
It was shown that the complexity of $n$ copies of a state \ket{\psi} is at most
\[QC(\ket{\psi}^{\otimes n}) \leq \log {n + 2^{QC(\ket{\psi})}-1 \choose 2^{QC(\ket{\psi})}-1}  + O(\log(n)) + O(\log(QC(\ket{\psi})) \]
The first term is the logarithm of the dimensions of the symmetric
subspace $H_{p}^{\otimes n}$ where $H_p$ is the space containing the
shortest description of \ket{\psi}, $p$. The second term comes from
describing $n$ in a self-delimiting way. The third term comes from
describing $\dim(H_p)=\log(QC(\ket{\psi})$ in a self-delimiting way.
The second and third terms contain a big-$O$ term since the
descriptions of $n$ and $\dim(H_p)$ are self-delimiting so that they
can be concatenated together with the symmetric vectors.

M\"{u}ller \cite{markuspaper,markusthesis} studied the precise definition of quantum
Kolmogorov complexity in detail, and gave a detailed analysis of the invariance theorem of
quantum Kolmogorov complexity. He showed that there is a universal reference Turing machine $U$
that can simulate another quantum Turing machine $T$ using a constant number
of qubits, even if $T$ and $U$ have access to a different finite sets of gates (given
that the time of computation is not known in advance). M\"{u}ller \cite{markus07} also gave a detailed
proof that classical strings have the same classical and quantum Kolmogorov complexities up to an additive constant factor.

\subsection{Quantum Kolmogorov Complexity based on Classical Descriptions}

Vitanyi \cite{vit01} defined a prefix quantum Kolmogorov complexity based on
classical descriptions.  The idea behind this complexity is as follows.
Suppose that \ket{p} is a self-delimiting classical input to a quantum Turing
machine $T$ which outputs a state $T\ket{p}=\ket{\phi}$ and that
\ket{\psi} is some state close to \ket{\phi}. It can be said that \ket{p} is a description
of \ket{\psi} but a penalty factor can be added depending on the distance between \ket{\psi}
and \ket{\phi}. Suppose $\ket{\psi}=\alpha
\ket{\phi}+\beta\ket{\phi^{\bot}}$  where $\ket{\phi^\bot}$ is
orthogonal to \ket{\phi}. Then the idea is that \ket{\phi} can be described by using
an additional $\log(|\beta^2|)$ to encode $\ket{\phi^{\bot}}$ using
lossless compression. Thus $\log(|\beta^2|)$ is the penalty factor depending on the distance
between \ket{\psi} and \ket{\phi}. The conditional prefix free string complexity
$K_Q$  of a string \ket{\psi} was defined as:
\[ K_Q (\ket{\psi}|y) = \min_{T(p,y)=\ket{\phi}} (l(p)+\lceil -\log(|\langle \psi | \phi \rangle |^2) \rceil)
\]
The first term $l(p)$ is the length of the shortest program, the second term is the number of bits used to encode
the orthogonal state to $\ket{\phi}$.

This complexity can be understood by considering an example.

\begin{example}[Quantum Kolmogorov complexity base on classical descriptions]
Consider describing the state
\[
\ket{\psi_n}=\frac{1}{\sqrt{2}} \ket{0} + \frac{1}{\sqrt{2}}\ket{n}
\]
as \ket{0}. Then the complexity of \ket{\psi_n} is
\be
K_Q(\ket{\psi_n}) &=& \log(c)  - \log( 1/2) \\
 &=& k
\ee
where $c$ and $k$ are constants independent of $n$.
\end{example}

More generally, any state
\[
\ket{\psi}=\sum_i \alpha_i \ket{i}
\]
can be described by describing the easiest part \ket{i} of \ket{\psi} and incurring a $-\log(1-|\alpha_i|^2)$ penalty
so that the complexity (based on classical descriptions) of \ket{\psi} is
bounded by
\[
K_Q(\ket{\psi}) \leq \min_i(K(i)+\log(1-|\alpha_i|^2))
\]

\subsection{Application of Classical Kolmogorov Complexity to Quantum Computation}

There have been several papers \cite{laplante-2005,laplante-2003,mor05,mor04,mor06,svo96} which apply classical Kolmogorov complexity directly to quantum mechanics. Laplante and Magniez \cite{laplante-2003} used Kolmogorov complexity to study lower bounds for randomized and quantum query complexity.  Laplante, Lee and Szegedy \cite{laplante-2005} also used classical Kolmogorov complexity to study quantum algorithms.
They produced new results in quantum computational complexity.
Laplante, Lee and Szegedy showed that the quantum adversary argument can be used on formula size lower bounds. Thus using results in quantum computing, they produced new results in classical computational complexity.

Svozil \cite{svo96} was the first to discuss defining a quantum Kolmogorov complexity back in 1996. Svozil used
a circuit based model of quantum computation,
and said that descriptions must be classical, so that the prefix complexity obeys Kraft's inequality. In the sense that the descriptions are purely classical, this is an application of
classical Kolmogorov complexity to quantum information theory.
Svozil described the basic properties of this complexity.

Mora and Breigel \cite{mor05,mor04} also used classical Kolmogorov complexity to define the complexity of quantum states.
According to Mora and Briegel, an
important feature of a description of a quantum object is that the description can be copied \cite{mor05}, therefore a quantum description is necessarily not quantum, but classical.
Therefore Mora and Briegel \cite{mor05,mor04}
defined the algorithmic entropy
of a quantum state to be the classical Kolmogorov
complexity of a circuit that describes the quantum state
up to a given degree $\epsilon$ (unfortunately this complexity increases beyond a constant factor as $\epsilon$ decreases).
Mora and Briegel showed that
the algorithmic complexity of a quantum state can be bounded in terms of the Schmidt number of its entanglement (although the converse does not hold).
Mora, Briegel and Kraus \cite{mor06} applied algorithmic complexity to various problems in quantum communication and computation
to produce new proofs of already known results.
An application of a quantum Kolmogorov complexity would be
to analyse a fully generalised quantum Maxwell's demon \cite{kolmcomplexity}. Mora, Briegel and Kraus \cite{mor06} also
applied the algorithmic complexity of quantum states
to begin to look at the effect of entanglement on Maxwell's demon.

\subsection{A Priori Probability}

Gacs \cite{gacs} and Tadaki \cite{tad02,tad04} both defined quantum Kolmogorov complexity like measures
based on probability theory. Gacs attempted to define a universal probability measure, which unfortunately
does not correspond to any length measure \cite{vit01}. Tadaki \cite{tad04} extended the semi-POVM to infinite dimensions and derived a quantum generalisation of Chaitin's halting probability \cite{chaitin} as the probability of any measurement outcome on the maximal infinite dimensional
semi-POVM.

\section{Second Quantised Kolmogorov Complexity}

There have been many attempts to define a quantum analogy of classical Kolmogorov complexity \cite{ber01,gacs,mor05,segre,svo96,tad02,tad04,vit01}.
In this paper, a modification of the scheme proposed
 in Berthiaume, van Dam and Laplante \cite{ber01} is taken \cite{kolmcomplexity,cazphdthesis}.

 As is usual in quantum information theory,
 Berthiaume {\it et al} assume that quantum strings are not in
superpositions of different lengths. However, a quantum
string might be in a superposition of a very simple string and a very complex string,
in which case, its ``shortest description" seems to naturally be a superposition of
a very short string and a very long string. Therefore it seems natural to allow
the shortest description to exist in a superposition of different lengths. The length
of the shortest description is defined to be its average length.

Second quantised Kolmogorov complexity can be defined with respect to a specific Turing machine $T$.
\begin{definition}[Second quantised Kolmogorov complexity with respect to $T$]
The second quantised Kolmogorov complexity $\Cl_T$ of a string \ket{\psi}
with respect to a quantum Turing machine $T$ is:
\[\Cl_T(\ket{\psi}) = \min_{T(\ket{\phi})=\ket{\psi}}\lbar(\ket{\phi}) \]
\end{definition}
Each quantum Turing machine has a finite number of instructions. The quantum Turing machines can therefore
be enumerated and a reference universal Turing machine can be defined using this enumeration.
In this definition, $\ket{1^{l(i)}0i}$ is simply a prefix free description of $i$.
\begin{definition}[Reference universal quantum Turing machine]
Let $T_1$, $T_2$, $\ldots$ be a enumeration of the quantum Turing machines. On input
$(i,\ket{\psi})=\ket{1^{l(i)}0i}\otimes \ket{\psi}$,
the reference universal quantum Turing machine
 simulates $T_i$ on input $\ket{\psi}$.
\end{definition}
The general second quantised complexity can be defined with respect to this reference universal quantum Turing machine.
\begin{definition}[Second quantised Kolmogorov complexity]
The second quantised Kolmogorov complexity $\Cl$ of a string \ket{\psi}
is defined the minimum average length of a description of \ket{\psi}
with respect to the reference quantum Turing machine $U$.
\[\Cl(\ket{\psi}) = \min_{U(\ket{\phi})=\ket{\psi}}\lbar(\ket{\phi}) \]

The conditional second quantised Kolmogorov complexity $C$ of a string \ket{\psi} given a string \ket{\chi}
is:
\[\Cl(\ket{\psi}| \ket{\chi}) = \min_{U(\ket{\chi},\ket{\phi})=\ket{\psi}}\lbar(\ket{\phi}) \]
\end{definition}

\subsubsection{Invariance}
Following the same reasoning for classical Kolmogorov complexity, second quantised
Kolmogorov complexity is invariant up to an additive constant
(for simplicity, it is assumed that all quantum Turing machines
have access to the same set of universal gates).
\begin{lemma}[Invariance]
Let $T$ be any quantum Turing machine. Then there exists a constant $c_T$, dependent
only on $T$, such that for any state \ket{\psi},
\[
\Cl(\ket{\psi})\leq \Cl_{T}(\ket{\psi})+c_T
\]
\end{lemma}

\subsubsection{Examples}

A simple example of second quantised Kolmogorov complexity is the complexity of
\[\ket{\phi_n} = \alpha\ket{0}+\beta\ket{n}\]
which is given by the average complexity of the two parts, \ket{0} and \ket{n}:
\[\Cl(\ket{\phi_n}) \leq |\beta|^2 \lceil \log(n)\rceil +c\]
Unlike Berthiaume {\it et al}'s quantum Kolmogorov complexity, the second quantised complexity of \ket{\phi_n} is
continuous at $|\beta|\rightarrow 0$

Another simple example of second quantised Kolmogorov complexity is to bound the complexity of a state from above
by its average length, using the identity machine.
\begin{lemma}[Upper bound on complexity]
\label{upper}
There exists a constant $c$ such that for any string $\ket{\psi}$
\[
\Cl(\ket{\psi})\leq \lbar(\ket{\psi})+c
\]
\end{lemma}
{\bf Proof.}
Let $T^I$ be the identity quantum Turing machine which copies its input to the output tape.
Then there exists some $j$ such that $T^I=T_j$ where $T_1$, $T_2$, $\ldots$ is the enumeration
of quantum Turing machines used in the definition of the reference universal quantum Turing machine.
On input $1^{l(j)}0j\ket{\psi}$, the universal quantum Turing machine outputs \ket{\psi}. Thus
any state \ket{\psi} can be described using at most $\lbar(\ket{\psi})+2l(j)+1$ qubits.
\qed

\subsection{Second Quantised Kolmogorov Complexity as Energy}

Rallan and Vedral described a physical interpretation of the average length
of indeterminate length strings as energy in a system in the second quantization \cite{ral02}.
This gives second quantised Kolmogorov complexity an intuitive physical interpretation.

A Hilbert space $H^{\otimes n}$ can be
realised by a sequence of photons $\ket{\phi_1}\otimes \ldots
\otimes \ket{\phi_n}$ in which \ket{\phi_i} represents exactly one
photon with frequency $\omega_i$. The value of the qubit
\ket{\phi_i} is realised by the polarisation of its photon, either
horizontal \ket{0} or vertical \ket{1}. The absence of a photon at a particular frequency can
be represented by \ket{\#} which is orthogonal to \ket{0} and
\ket{1}. Indeterminate length strings are obtained by allowing the
number of photons to exist in superposition and ordering the photons by their frequencies.
The first \ket{\#} (which can be in a superposition of positions) is used to mark the end of the string.

The frequency of each
photon \ket{\phi_i} is chosen to be approximately equal so that $\omega_i \approx
\omega$ for some value $\omega$. The energy in a superposition of photons
is the average energy required to either create or destroy that
superposition ($\hbar \omega$ per photon of frequency $\omega$ where $\hbar$
is Planck's constant).
Thus the energy of an indeterminate length string of photons
\ket{\phi} is proportional to its average length and is given by (approximately)
$\hbar \omega \lbar(\ket{\phi})$. The average length of an indeterminate length
quantum string is therefore proportional to the energy within the string.

In this physical implementation of quantum mechanics, the second quantised Kolmogorov complexity
of a state represents the minimum energy required to describe that state.

\subsection{Incompressibility}

A fundamental property of classical Kolmogorov complexity is that it is incompressible.
Second quantised Kolmogorov complexity is also incompressible in the sense that it is not
possible to compress a set of quantum strings arbitrarily.
This incompressibility theorem is analogous to the incompressibility theorem given by Berthiaume {\it et al} \cite{ber01}.
\begin{theorem}[Incompressibility]
\label{incomp}
Let  \ket{\psi_1}, $\ldots$, \ket{\psi_{m}} be quantum strings (which are possibly non-orthogonal).
Then there exists $1\leq i \leq m$ such that:
\[\Cl(\ket{\psi_i}) \geq  \frac{S(\rho)-1}{2}\]
where $\rho = \sum_{j=1}^m \frac{1}{m} \ketbra{\psi_i}$.
\end{theorem}
{\bf Proof.}
The result is an application Schumacher and Westmoreland's average length compression scheme \cite{sch01} (which was
described in Section \ref{lossyqcoding}). They showed that given a mixture of states
\[
\mixture = \{(1/m, \ket{\psi_i})\}_{i=1}^m
\]
if $U$ is any prefix free encoding of the \ket{\psi_i}'s then the expected average length after compression
is bounded below by $\rho$'s von Neumann entropy:
\be
E(\lbar(U(\mixture))) &=& \sum_i \frac{1}{m}\lbar(U(\ket{\psi_i})) \\
&\geq & S(\rho)
\ee
(where $\rho$ is defined as in the theorem). Since $E(\lbar(U(\mixture)))\geq S(\rho)$, it must be the case that
for some $i$,
\[
\lbar(U(\ket{\psi_i})) \geq S(\rho) \label{geq}
\]
So far, it has been assumed that $U$ is prefix free. Suppose that \mixture is compressed using a variable length
unitary code $V$ (which is not necessarily prefix free). Let $S$ be the unitary operation
\[
S \ket{x} = \ket{1^{l(x)}0x}
\]
so that $S$ transforms each string \ket{x} into a self-delimiting string $S\ket{x}$. In which case, $SV$ is a
prefix free mapping. Using Eq. \ref{geq}, there is some \ket{\psi_i} such that:
\be
2 \lbar(V(\ket{\psi_i})) +1 &=& \lbar(SV(\ket{\psi_i})) \label{incompbound} \\
&\geq& S(\rho)
\ee
Rearranging gives the result.
\qed

The bound in the theorem can of course be improved by choosing a more efficient
encoding of the length information in the proof (i.e. choosing a different
function $S$).

\subsection{Complexity of Multiple Copies}

A simple but non-trivial example of second quantised Kolmogorov complexity is the complexity of $n$ copies of a qubit.
The second quantised complexity of $n$ copies can be much smaller than the corresponding quantum Kolmogorov complexity
of Berthiaume {\it et al} \cite{ber01}.

\begin{theorem}[Complexity of Multiple Copies]
Let \[\ket{\psi}=\alpha \ket{0}+\beta\ket{1}\]
Let $X$ be a random variable such that for $0\leq i \leq n$:
\[P(X=i)= |\alpha|^{2i}|\beta|^{2(n-i)}\]
Then, given $n$, the conditional complexity of $\ket{\psi}^{\otimes n}$ is at most:
\[
C(\ket{\psi^{\otimes n}}|\ket{n}) \leq H(X)+c_X
\]
where $c_X$ is dependent on $X$ but not on $n$.
\end{theorem}
{\bf Proof.}
$n$ copies of $\ket{\psi}$ can be expanded out as:
\be
\ket{\psi^{\otimes n}}&=& (\alpha\ket{0}+\beta\ket{1})^{\otimes n}\\
&=&\sum_{i=0}^{n}\alpha^i \beta^{n-i} S(i,n)
\ee
where $S(i,n)$ is the symmetric superposition of strings containing $i$
\ket{0}'s and $n-i$ \ket{1}'s. Given $n$ and a description of $X$,
 each $S(i,n)$ can be described as a string of length
$\log(|\alpha^{i} \beta^{n-i}|^2)$ qubits. Thus each $\ket{\psi^{\otimes n}}$ can be described using
$H(X)+c_X$ qubits, where the constant $c_X$ comes from describing
the probability distribution $X$.
\qed

This bound is not tight.
If some state \ket{\psi} is close to \ket{+}, then $H(X)$ is large however if a basis
change is performed (using $n$ Hadamard operations) from $\{\ket{0}, \ket{1}\}$ to $\{\ket{-}, \ket{+}\}$,
then $H(X)$ is dramatically reduced in this basis.

\section{Prefix Complexity}

A prefix Kolmogorov complexity can be defined by defining a prefix quantum Turing machine which
takes as input prefix free strings.
\begin{definition}[Prefix quantum Turing machine]
A quantum Turing machine $T$ is a prefix quantum Turing machine if any two orthogonal inputs
to $T$ are prefix free.
\end{definition}
As with classical Kolmogorov complexity, all inputs can be defined to have infinite length, and the average of the maximum lengths of the number of qubits used (on the input tape) during computation can be defined to be the length of the input. The prefix quantum Turing machines can then be enumerated to define a reference universal prefix
quantum Turing machine in the same manner as for classical Kolmogorov complexity \cite{vitanyi}.
A prefix
complexity can be defined with respect to the reference prefix Turing machine.
\begin{definition}[Second quantised prefix Kolmogorov complexity]
Let $U$ be the universal reference prefix quantum Turing machine. The second quantised prefix Kolmogorov complexity $\Kl$
of a state $\ket{\psi}$ is:
\[ \Kl(\ket{\psi})=\min_{U\ket{\phi}=\ket{\psi}} \lbar(\ket{\phi})
\]
\end{definition}

\subsection{Incompressibility}

In proving the incompressibility theorem for the non-prefix complexity (Theorem \ref{incomp}) the incompressibility theorem for prefix complexity has inadvertently been proved. Here it is.
\begin{theorem}[Incompressibility of prefix complexity]
\label{qKincomp}
Let  \ket{\psi_1}, $\ldots$, \ket{\psi_{m}} be quantum strings (which are possibly non-orthogonal).
Then there exists $1\leq i \leq m$ such that:
\[\Kl(\ket{\psi_i}) \geq  S(\rho)\]
where $\rho = \sum_{j=1}^m \frac{1}{m} \ketbra{\psi_i}$.
\end{theorem}
{\bf Proof.}
See proof of Theorem \ref{incomp}.
\qed

\subsection{Relationship of Prefix Complexity to von Neumann entropy}

Since the expected average length of an optimal encoding of a mixture is its von Neumann entropy,
prefix complexity is related to von Neumann entropy.
Let us define the expected complexity of a density
operator in the same way that it is defined classically.
\begin{definition}[Expected prefix Kolmogorov complexity of $\rho$]
Let
\[
\rho = \sum_i p_i \ketbra{\psi_i}
\]
be a density operator in its diagonalised form. The expected prefix Kolmogorov complexity of $\rho$ is:
\[
E(\Kl(\rho)) = \sum_i p_i \Kl(\ketbra{\psi_i})
\]
\end{definition}
A density operator may have more than one possible diagonalisation. The complexity of a density operator is therefore defined to be the minimum complexity of describing a mixture that corresponds
to that density operator.
\begin{definition}[Prefix Kolmogorov complexity of $\rho$]
Let $\mixture=\{(p_i,\ket{\psi_i})\}$ be a mixture of quantum states.
Let $\leq$ be a total ordering on quantum states and without loss of generality,
suppose that $\ket{\psi_1}\leq \ket{\psi_2} \leq \ldots \leq \ket{\psi_m}$. An encoding
$f$ of \mixture is:
\[
f(\mixture) = ((p_1,\ket{\psi_1}), \ldots, (p_m,\ket{\psi_m}))
\]
The second quantised prefix Kolmogorov complexity of \mixture is defined as:
\[
\Kl(\mixture) = \Kl(f(\mixture))
\]
The second quantised prefix Kolmogorov complexity of a density operator $\rho$ is defined as:
\[
\Kl(\mixture) = \min_{\mixture=\{(p_i,\ket{\psi_i}\}_i \txt{ and } \sum_i p_i\ketbra{\psi_i}=\rho } \Kl(f(\mixture))
\]
\end{definition}
The relationship between von Neumann
is the same as for Shannon entropy and classical prefix complexity.
\begin{theorem}[Relationship of $S$ and $\Kl$]
Let $\mixturedef$ be a mixture of quantum states with density operator $\rho$.
Let
\[E(\Kl(\mixture))=\sum_i p_i \Kl(\ket{\psi_i}) \]
Then there exists a constant $c$, independent of $\rho$, such that
\[ E(\Kl(\rho))-\Kl(\rho)-c \leq S(\rho) \leq E(\Kl(\rho)) \]
\end{theorem}
{\bf Proof.}
By Schumacher and Westmoreland's
result \cite{sch01}, if \mixture is encoded by a prefix free encoding, then the expected average length
of that prefix free encoding is bounded below by the von Neumann entropy of the mixture's density operator.
Since the orthogonal inputs to a prefix Turing machine are prefix free, their expected length
is bounded below by $S(\rho)$. On the other hand, an expected length of $S(\rho)+1$ qubits can be achieved
by encoding $\rho$ in its diagonal basis. There exists a Turing machine $T_\rho$ (which can be described using
approximately $\Kl(\rho)$ qubits) which carries out
this encoding, hence the result follows.
\qed

\subsection{Additivity of Superpositions}

The incompressibility theorem (Theorem \ref{qKincomp}) can
also be used to show that prefix second quantised Kolmogorov complexity is not additive in the sense that the complexity of a superposition is not bounded above or below by the complexity of its parts. In other words, the
following two inequalities do not hold.
\be
\Kl\left( \sum_i \alpha_i \ket{\psi_i} \right) &\leq & \sum_i |\alpha_i|^2 \Kl(\ket{\psi_i} )+c \\
\Kl\left( \sum_i \alpha_i \ket{\psi_i} \right) &\geq & \sum_i |\alpha_i|^2 \Kl(\ket{\psi_i} )-c
\ee
This additivity
does not hold even when the states in the superposition are restricted to being orthogonal.
\begin{theorem}[Non-additivity of second quantised complexity]
For any constant $k>0$, there exist orthogonal states \ket{\psi_1}, \ket{\psi_2}, $\ldots$, \ket{\psi_n} and constants
$\alpha_1$, $\ldots$, $\alpha_n$ such that
\[
\Kl\left( \sum_i \alpha_i \ket{\psi_i} \right) > \sum_i |\alpha_i|^2 \Kl(\ket{\psi_i} ) + k
\]
For any constant $k>0$, there exist orthogonal states \ket{\psi_1}, \ket{\psi_2}, $\ldots$, \ket{\psi_n} and constants $\alpha_1$, $\ldots$, $\alpha_n$ such that
\[
\Kl\left( \sum_i \alpha_i \ket{\psi_i} \right) <  \sum_i |\alpha_i|^2 \Kl(\ket{\psi_i} ) - k
\]

\end{theorem}
{\bf Proof.}
For $n\geq 0$, let
\be
\ket{\phi_{n}^+} &=& \frac{\ket{0}+\ket{n}}{\sqrt{2}} \\
\ket{\phi_{n}^-} &=& \frac{\ket{0}-\ket{n}}{\sqrt{2}}
\ee
\\
($>$)
For each $m$, consider the states
\ket{2^m}, $\ldots$, $\ket{2^{m+1}-1}$.
By the incompressibility theorem above, there exists some $2^m\leq n < 2^{m+1}$ such that
\be
\Kl(\ket{n}) &\geq& S\left(\sum_{i=2^m}^{2^{m+1}-1} \frac{\ketbra{i}}{2^m} \right)  \\
&=& m \\
&=& \lfloor \log(n) \rfloor
\ee
On the other hand
\[
\ket{n} = \frac{\ket{\phi_{n}^+} - \ket{\phi_{n}^-}}{\sqrt{2}}
\]
But by Lemma \ref{upper},
\be
\frac{1}{2}\Kl(\ket{\phi_{n}^+})+\frac{1}{2}\Kl(\ket{\phi_{n}^-}) &\leq & \frac{\lbar(\ket{\phi_{n}^+}) + \lbar(\ket{\phi_{n}^-})}{2}+c\\
&=&\frac{\log(n)}{2}+c
\ee
Thus for any constant $k$, there exists sufficiently large $n$ such that:
\be
\Kl\left(\frac{\ket{\phi_{n}^+} - \ket{\phi_{n}^-}}{\sqrt{2}}\right) &=&\Kl(\ket{n}) \\
&\geq& \lfloor \log(n) \rfloor \\
&> &\frac{\log(n)}{2}+c+k\\
&\geq&\frac{\Kl\left(\ket{\phi_{n}^+}) + \Kl(\ket{\phi_{n}^-}\right)}{2}+k
\ee
\\
($<$)
By Lemma \ref{upper},
\[
\Kl(\ket{0}) \leq c
\]
On the other hand, there exists sufficiently large $N$ such that the density operator
\[
\rho = \sum_{i=1}^{N} \frac{\ketbra{\phi_{i}^+}}{N}
\]
has arbitrarily high entropy $S(\rho)$. Therefore, by the incompressibility theorem, for any constants $c$ and $k$
there exists $n$ such that $\Kl(\ket{\phi_{n}^+})>2(c+k)$. For this $n$:
\be
\Kl(\ket{0}) &\leq & c\\
&<& \frac{\Kl(\ket{\phi_{n}^+})}{2} -k\\
&<& \frac{\Kl(\ket{\phi_{n}^+})+\Kl(\ket{\phi_{n}^-})}{2}-k
\ee
which completes the proof.
\qed

\subsection{Inequalities of Von Neumann Entropy}

It has been
shown that the same inequalities hold for classical Kolmogorov complexity
and Shannon entropy \cite{kineqs} and
that the same inequalities hold for classical Kolmogorov complexity
and the size of sets \cite{astrangeapp,combinatorialapproach}.
However, very little is known about the inequalities for von Neumann entropy and most of the known inequalities for von Neumann entropy can be derived
from subadditivity \cite{anewineq}. We now show that
the inequalities obeyed by second quantised
prefix complexity are also obeyed by von Neumann entropy.

To discuss general inequalities, we first define some
notation. For this, we follow Hammer, Romashchenko and Vereshchagin \cite{kineqs}.
Let $\rho^{X_1\otimes \ldots \otimes X_n}=\rho^{X_1}\otimes
\ldots\otimes\rho^{X_n}$
be any $n$-partite state state in the Hilbert space
$X_1\otimes \ldots \otimes X_n$.
For any $W\subseteq \{1,\ldots, n\}$, let
\[
\rho^{W_X} = \rho^{\bigotimes_{i\in W} X_i}
\]
Then, since conditional and mutual entropies can
be expanded out as the sums of joint entropies, any inequality of von Neumann entropy can be
written as
\[
\sum_{W\in Z}\lambda_W S(\rho^{W_X}) \geq 0
\]
for some $n$, for some Hilbert space $\rho^{X_1}\otimes
\ldots\otimes\rho^{X_n}$ and
for some set $Z$ of subsets of $\{1,\ldots,n\}$
where the $\lambda_W$'s are real constants.
\begin{theorem}[Inequalities of von Neumann entropy]
Let $Z$ be a set of subsets of $\{1,\ldots,n\}$.
For each subset $W$ of $\{1,\ldots,n\}$, let
$\lambda_W$ be some real constant. Then if:
\[
\sum_{W\in Z}\lambda_W E(\Kl(\rho^{W_X})) \geq 0 \label{ineq1}
\]
holds for all $n$-partite spaces $X_1\otimes \ldots \otimes X_n$
and for all states $\rho \in X_1\otimes \ldots \otimes X_n$, then the inequality
\[
\sum_{W\in Z}\lambda_W S(\rho^{W_X}) \geq 0
\]
also holds for all density operators $\rho$ in all $n$-partite spaces $X_1\otimes \ldots \otimes X_n$.
\end{theorem}
{\bf Proof.}
Let $\rho$ be any density operator in some $n$-partite
space $X_1\otimes \ldots \otimes X_n$.
Then for each set $W\subseteq \{1,\ldots,n\}$ and each integer $m>0$:
\be
mS(\rho^{W_X}) &=& S((\rho^{W_X})^{\otimes m}) \\
&=& S\left( \bigotimes_{i\in W} (\rho^{X_{i}})^{\otimes m} \right)
\ee
Let $Y_{i}^m = X_{i}^{\otimes m}$ (we shall apply
Eq. \ref{ineq1} to $W_{Y^m}$ rather than $W_X$). Then
\be
mS(\rho^{W_X}) &=& S\left( \bigotimes_{i\in W} (\rho^{X_{i}})^{\otimes m} \right) \\
&=& S\left( \bigotimes_{i\in W} \rho^{Y_{i}^m} \right) \\
&=& S( (\rho^{\otimes m})^{W_{Y^m}})
\ee
Now $(\rho^{\otimes m})^{W_{Y^m}}$ can be described by describing $m$, $\rho$ and $W$. Thus
\be
S(\rho^{W_X})&=&\frac{S( (\rho^{\otimes m})^{W_{Y^m}})}{m}\\
&=& \frac{E(\Kl((\rho^{\otimes m})^{W_{Y^{m}}})) +O(\log(m))
+c_\rho+O(n)}{m}
\ee
And so, as $m$ grows large,
\be
\sum_{W\in Z}\lambda_W S(\rho^{W_X})
&=& \sum_{W\in Z} \lambda_W \frac{E(\Kl((\rho^{\otimes m})^{W_{Y^{m}}})) +O(\log(m)) +c_\rho+O(n)}{m} \nonumber\\
&\rightarrow&  \sum_{W\in Z} \lambda_W \frac{\Kl((\rho^{\otimes m})^{W_{Y^{m}}})}{m} \\
\ee
Assuming that the corresponding inequality holds for second quantised complexity for all states $\rho$ in all $n$-partite spaces  (Eq. \ref{ineq1}), as $m$ grows large
\be
\sum_{W\in Z}\lambda_W S(\rho^{W_X})
&\leftarrow&  \sum_{W\in Z} \lambda_W \frac{E(\Kl((\rho^{\otimes m})^{W_{Y^{m}}}))}{m} \\
&\geq & 0 \nonumber
\ee
\qed

\section{Conclusions}

We have defined a second quantised Kolmogorov complexity that has an intuitive physical interpretation.
We have discussed its basic properties. We have also shown that
the second quantised prefix complexity is closely
related to von Neumann entropy.

\section{Acknowledgments}
We thank Sougato Bose and Andreas Winter for useful comments. We thank Markus
M\"{u}ller for making valuable points on previous versions of this work. We thank the Engineering and Physical Sciences Research
Council in the UK for financial support. V. Vedral thanks the
Royal Society and Wolfson Foundation
for financial support. This work was partly supported by the National Research Foundation and Ministry of Education (Singapore).

\bibliographystyle{plain}
\bibliography{AverageLengthComplexity}

\end{document}